\documentclass[prl,twocolumn,superscriptaddress,floatfix,letterpaper]{revtex4}
\pdfoutput=1

\newcommand{\vect}[1]{\mathbf{#1}}

\begin{document}

\title{
High temperature fractional quantum Hall states
}

\author{Evelyn Tang}
\affiliation{Department of Physics, Massachusetts Institute of Technology, Cambridge, Massachusetts 02139, USA}
\author{Jia-Wei Mei}
\affiliation{Department of Physics, Massachusetts Institute of Technology, Cambridge, Massachusetts 02139, USA}
\affiliation{Institute for Advanced Study, Tsinghua
University, Beijing, 100084, P. R. China}
\author{Xiao-Gang Wen}
\affiliation{Department of Physics, Massachusetts Institute of Technology, Cambridge, Massachusetts 02139, USA}

\date{Nov, 2010}
\begin{abstract}
We show that a suitable combination of geometric frustration, ferromagnetism
and spin-orbit interactions can give rise to nearly flat bands with a large
bandgap and non-zero Chern number.  Partial filling of the flat band can give
rise to fractional quantum Hall states at high temperatures (maybe even room
temperature).  While the identification of material candidates with suitable
parameters remains open, our work indicates intriguing directions for exploration and synthesis. 
\end{abstract}

\maketitle

Landau symmetry breaking\cite{L3726,LanL58} has been the standard theoretical
concept in the classification of phases and transitions between them. However,
this theory turned out insufficient when the fractional quantum Hall (FQH)
state\cite{TSG8259,L8395} was discovered. These highly-entangled quantum states
are not distinguished by their symmetries; instead they are characterized by
new topological quantum numbers such as robust ground state
degeneracy\cite{Wtop,WNtop} and robust non-Abelian Berry's phases\cite{WZ8411}
of the ground states\cite{Wrig}. The new kind of order revealed in these
topological quantum numbers is named topological order.\cite{Wrig,Wtoprev}
Recently it was realized that topological order can be interpreted as patterns
of long-range quantum entanglement\cite{KP0604,LW0605,CGW1035}. This
entanglement has important applications for topological quantum computation:
the robust ground state degeneracy can be used as quantum memory\cite{DKL0252}.
Fractional defects from the entangled states which carry fractional
charges\cite{L8395} and fractional statistics\cite{LM7701,W8257,ASW8422} (or
non-Abelian statistics\cite{MR9162,Wnab}) can be used to perform fault tolerant
quantum computation\cite{K032,NSS0883}.

At present, the highly entangled gapped phases in FQH
systems\cite{TSG8259,L8395} are only realized at very low temperatures. In this
paper, we present a proposal to realize these states at high temperatures (even
room temperature).  The ideal is to combine spin-orbit coupling,
ferromagnetism, and geometric frustration.  Both spin-orbit coupling and
ferromagnetism can have high energy scales and can appear at room temperature.
In some cases, combining them leads to energy bands with non-zero Chern numbers
and filling such an energy band will give rise to integer quantum Hall states.
Further, in geometrically frustrated systems --- lattices on which hopping is
frustrated --- some of these topologically non-trivial energy bands can be very
flat\cite{Katsura,Chamon}. These would mimic Landau levels in free space. When
such a flat band with a non-zero Chern number is partially filled (such as 1/3 or
1/2 filled), FQH states can appear. Here we study a simple example of this idea
on the geometrically frustrated kagome lattice.

Several aspects of the above ideas have been studied actively in recent
research. Spin-orbit coupling can lead to a topological insulator in various
geometrically frustrated systems,\cite{LZW0923,GF0902,GF0905,WZ0965} and
non-collinear magnetic order can lead to integer quantum Hall
states.\cite{OMN0065,GF0902} Alternatively, interactions in geometrically
frustrated systems can break time-reversal
symmetry\cite{KTS0207,TU0304,PFS0804,WRW1025,RH1006,HZW1068,LYM1002,NNO1008,Martin}
which again can give rise to integer quantum Hall states. Here we will show
that a natural extension of these ideas may set the stage for FQH states and
other highly entangled states with fractional statistics and fractional charges
--- possibly even at room temperature.

\begin{figure}
\includegraphics[scale=0.158]{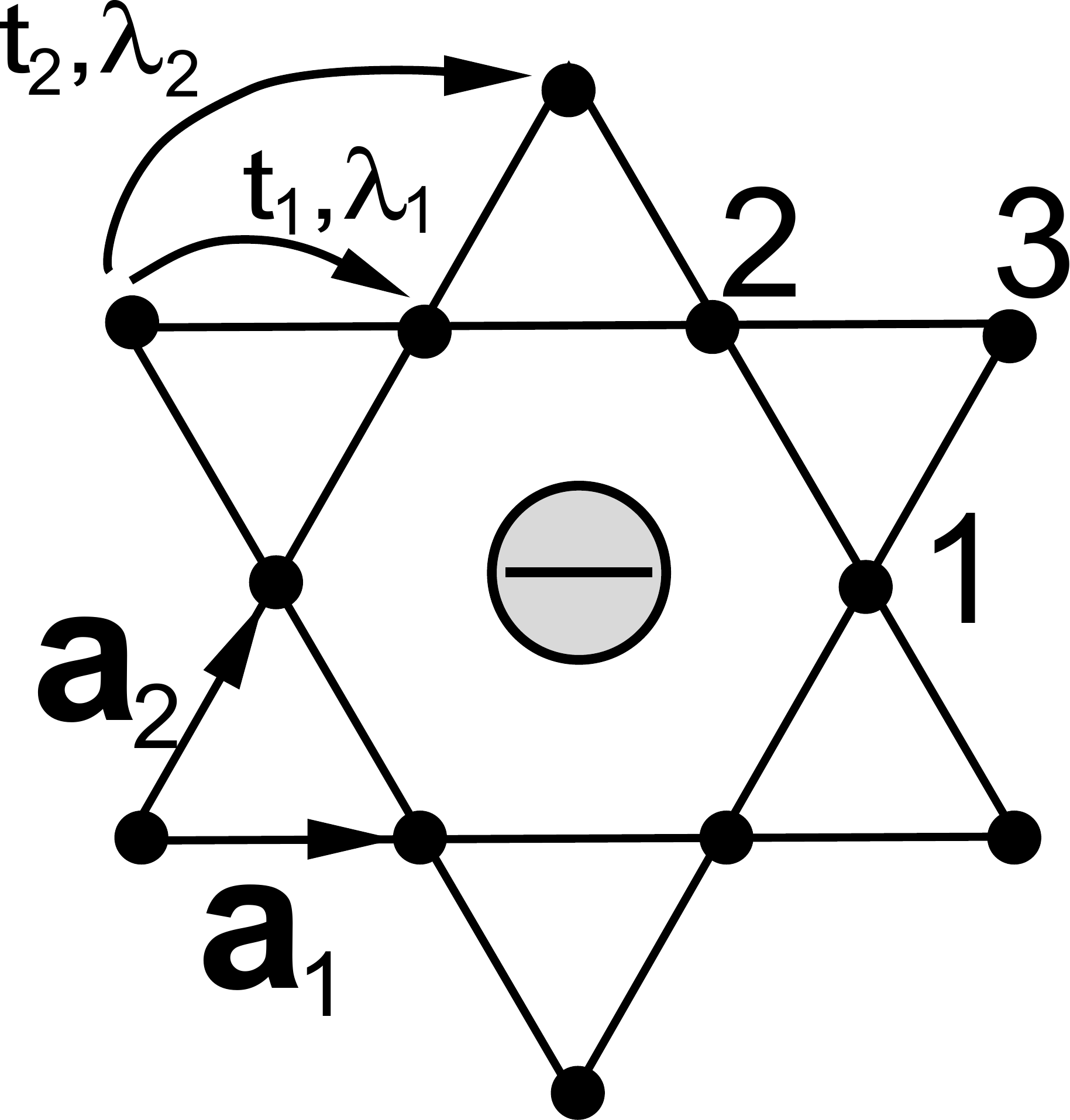}
\caption{The kagome lattice is a triangular Bravais lattice with a 3-point
basis labelled $l=1,2,3$; $\vect{a}_1=\hat{x}$ and
$\vect{a}_2=(\hat{x}+\sqrt{3}\hat{y})/2$ are the basis vectors. In the metallic
kagome lattice $Fe_3Sn_2$, spin-orbit coupling arises from the electric field
due to the Sn ion at the center of the hexagon.} 
\label{latticepic}
\end{figure}

\emph{Nearly flat band with non-zero Chern number}
We consider nearest and next-nearest neighbor hopping on a kagome lattice with
spin-orbit interactions described by the following Hamiltonian
\begin{eqnarray}
&&H=-t_1\sum_{\langle ij\rangle\sigma}c^\dag_{i\sigma}c_{j\sigma}+i\lambda_1\sum_{\langle ij\rangle\alpha\beta}(\vect{E}_{ij} \times \vect{R}_{ij})\cdotp\boldsymbol\sigma_{\alpha\beta} c^\dag_{i\alpha}c_{j\beta}\nonumber\\
&&-t_2\sum_{\langle\langle ij\rangle\rangle\sigma}c^\dag_{i\sigma}c_{j\sigma}+i\lambda_2\sum_{\langle\langle ij\rangle\rangle\alpha\beta}(\vect{E}_{ij} \times \vect{R}_{ij})\cdotp\boldsymbol\sigma_{\alpha\beta} c^\dag_{i\alpha}c_{j\beta}
\label{ham}
\end{eqnarray}
where $c^\dag_{i\sigma}$ creates an electron with spin $\sigma$ on the site
$\vect{r}_i$ on the kagome lattice. Here $\langle ij\rangle$ denotes nearest
neighbors and $\langle\langle ij\rangle\rangle$ next-nearest neighbors. The
second and fourth terms describe spin-orbit interactions which preserve
time-reversal invariance. 
$\vect{R}_{ij}$ is the distance vector between sites $i$ and $j$ and
$\vect{E}_{ij}$ the electric field from neighboring ions experienced along
$\vect{R}_{ij}$. 

To obtain FQH states, we need to break time-reversal symmetry. This is likely
to happen spontaneously from exchange effects in the flat band that cause
ferromagnetism\cite{KTS0207,TU0304,PFS0804}. Alternatively, one can apply an
external magnetic field or couple the system to a ferromagnet. In the extreme
limit the electron spins are totally polarized within the partially filled
band, which is the case we examine here. Hence we consider spin-orbit coupling
that also conserves $S_z$, i.e. the electric field on each site is in the 2D
plane.

We first study just nearest-neighbor hopping so $t_2=\lambda_2=0$. In momentum space,
Eq. \ref{ham} becomes
\begin{eqnarray}
H_\vect{k}&=&-2t_1
\left(
\begin{array}{ccc}
0 & \cos k_1 &\cos k_2 \\
\cos k_1 & 0 & \cos k_3 \\
\cos k_2 & \cos k_3 & 0
\end{array}
\right)\nonumber\\
&\pm&i2\lambda_1\left(
\begin{array}{ccc}
0 & \cos k_1 &-\cos k_2 \\
-\cos k_1 & 0 & \cos k_3 \\
\cos k_2 & -\cos k_3 & 0
\end{array}
\right)
\end{eqnarray}
where $\vect{a}_1=\hat{x}$, $\vect{a}_2=(\hat{x}+\sqrt{3}\hat{y})/2$,
$\vect{a}_3=\vect{a}_2-\vect{a}_1$ and $k_n=\vect{k}\cdot\vect{a}_n$.  We use
units where the hopping parameter $t_1=1$. 
The $+(-)$ sign refers to spin up (down)
electrons; from here we focus on just the spin up electrons. 

The spectrum consists of three energy bands and is gapless at $\lambda_1=0,
\pm\sqrt{3}$. At all other points the spectrum is gapped and the top and bottom
bands have unit Chern number with opposite sign while the middle band has zero
Chern number.  Here the Chern number is defined as\cite{TKN8205} 
\begin{eqnarray}
c=\frac{1}{2\pi }\int_{BZ}d^2k F_{12}(k)\label{fs}
\end{eqnarray}
where $F_{12}(k)$ is the associated field strength given by
$F_{12}(k)=\frac{\partial}{\partial k_1}A_2(k)-\frac{\partial}{\partial
k_2}A_1(k)$ with the Berry connection $A_\mu(k)=-i\bra
{n_\vect{k}}\frac{\partial}{\partial k_\mu}\ket{n_\vect{k}}$.  In the above
$\ket{n_\vect{k}}$ is a normalized wave function of the respective band.

Focusing on the lowest band which has a non-zero Chern number, we look for a
regime in which this band is very flat compared to the bandgap and the energy
scale of interactions. We denote $W$ as the maximum bandwidth of the lowest
band, $\Delta_{12}$ as the minimum bandgap between the two lowest bands and $U$
as the strength of electron-electron interactions in our system. When $U\gg W$,
interaction effects dominate kinetic energy and partially filling the flat band
would favor the Laughlin state.\cite{L8395} Since band mixing could destroy the
band flatness, ideally $\Delta_{12}\gg U$.  Hence we aim to maximize the ratio
$\Delta_{12}/W$
in order to obtain FQH states.

As the middle band has zero Chern number, any mixing between only two of the
bands would not change the Chern number of the lowest band. If the lowest band
remains flat even with mixing then $\Delta_{13}$, the minimum bandgap between
the lowest and highest bands, (and consequently the ratio $\Delta_{13}/W$) is also of interest. 

\begin{figure}
\includegraphics[width=3.23in]{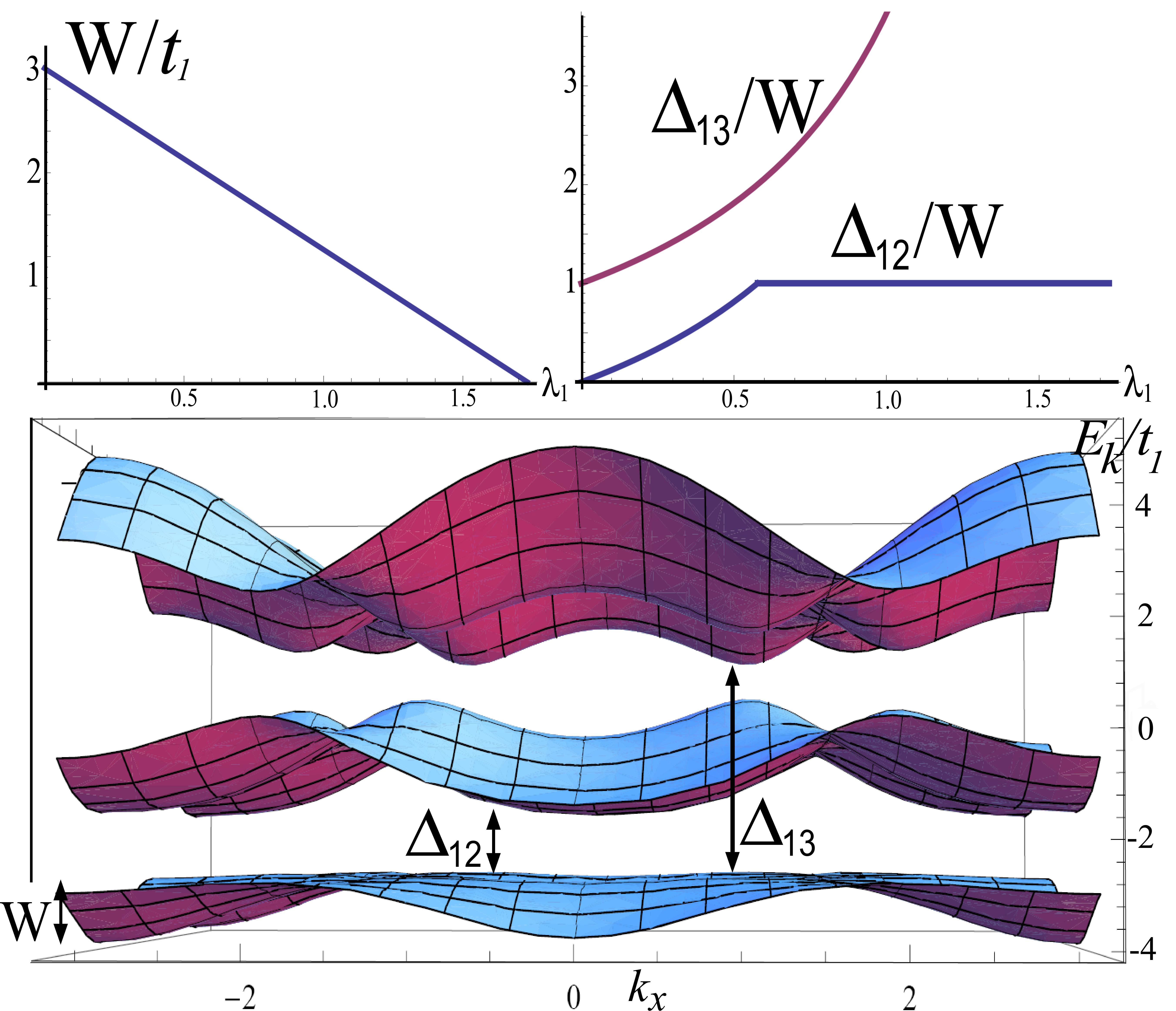}
\caption{(Color online) Results for nearest neighbor hopping as a function of
$\lambda_1$ (nearest neighbor spin-orbit coupling). The bandwidth of the lowest
band $W$ vanishes at $\lambda_1=\sqrt{3}$; however $W\geq\Delta_{12}$ always
where $\Delta_{12}$ is the bandgap between the two lowest bands. Here we show
the band structure for $\lambda_1=1$ where $W=1.3$ and $\Delta_{13}/W=3.7$. The
spectrum does not have a clear separation of energy scales between
$\Delta_{12}$, $W$ and $U$, the interaction strength, which could be due to
limitations of the nearest neighbor hopping model. }
\label{nn}
\end{figure}

We find that $W\geq\Delta_{12}$ always --- as the bandwidth vanishes, the
bandgap between the two lower bands vanishes in the same way (due to
topological symmetry in real-space)\cite{Bergman}, see Fig. \ref{nn}. Here we
show the band structure for $\lambda_1=1$ where $W=1.3$ and
$\Delta_{13}/W=3.7$. As $\Delta_{12}/W\leq 1$ always, the spectrum does not
have a clear separation of energy scales. When interactions are on the order of
$W$, the bands will mix. This scenario is quite different from Landau levels in
free space that are flat and well-separated, 
which could be due to limitations of this simplest model. 

For a more realistic scenario, next we include second-nearest neighbor hopping.
This gives us additional terms in the
Hamiltonian
\begin{eqnarray}
H_\vect{k}&=&-2t_2\left(
\begin{array}{ccc}
0 & \cos (k_2+ k_3) &\cos (k_3-k_1) \\
 & 0 & \cos (k_1+k_2) \\
 & & 0
\end{array}
\right)\nonumber\\
&+&i2\lambda_2\left(
\begin{array}{ccc}
0 & -\cos (k_2+ k_3) &\cos (k_3-k_1) \\
 & 0 & -\cos (k_1+k_2) \\
 & & 0
\end{array}
\right)
\end{eqnarray}

\begin{figure}
\begin{center}
\includegraphics[width=2.8in]{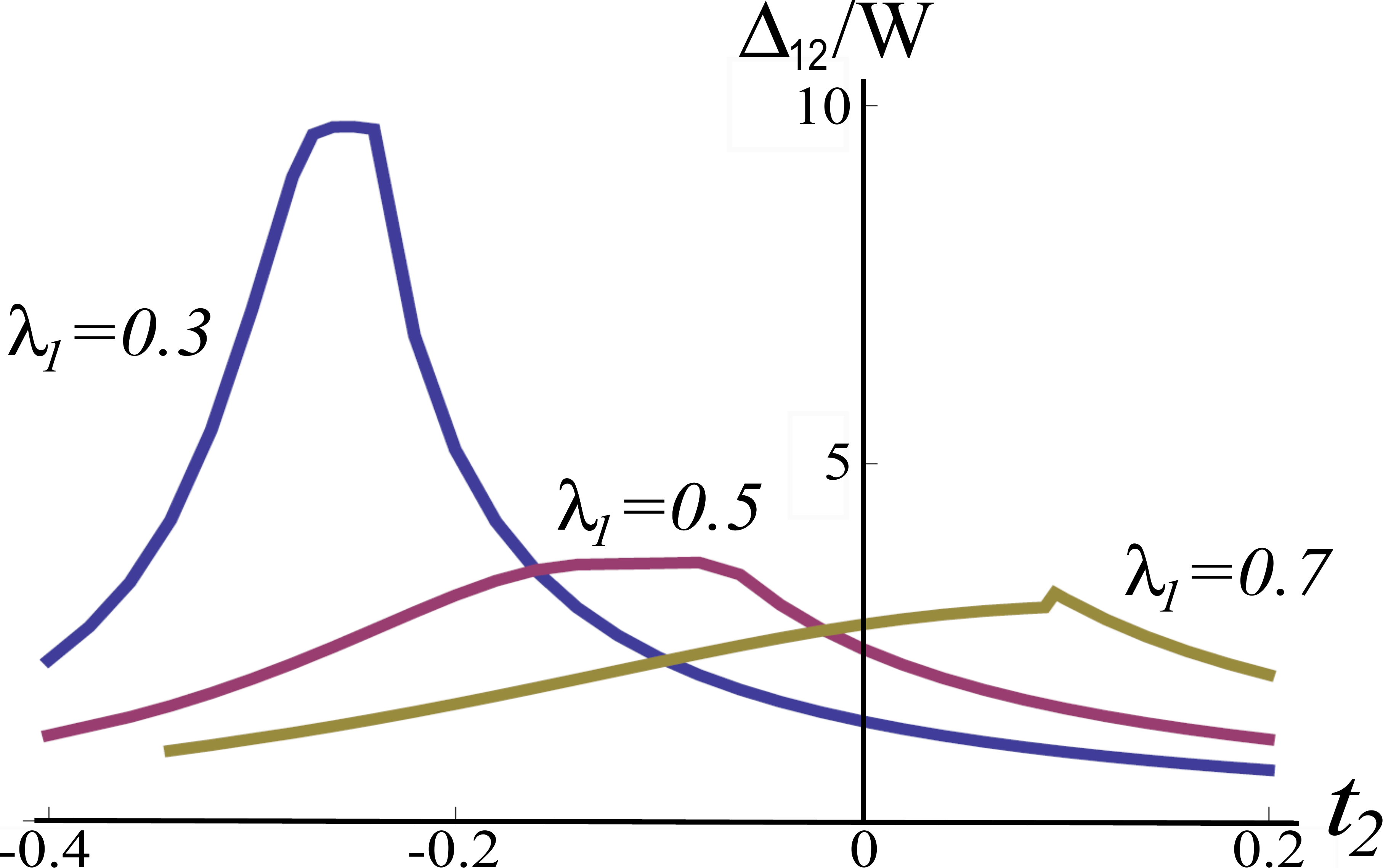}
\end{center}
\caption{(Color online) With the inclusion of next-nearest neighbor hopping, we
obtain much higher bandgap to bandwidth ratios. We choose three values of
$\lambda_1=\lambda_2=0.3, 0.5$ and $0.7$, and sweep $\Delta_{12}/W$ with $t_2$.
For lower values of spin-orbit coupling, the ratio peaks at negative $t_2$; for
relatively higher values of spin-orbit coupling the converse is true.} 
\label{cuts}
\end{figure}

In this larger parameter space the band maxima and minima are no longer fixed
at the same symmetry points.  We find that the largest values of
$\Delta_{12}/W$ (and $\Delta_{13}/W$) occur when $\lambda_1$ and $\lambda_2$
are of the same sign -- in which case the results are symmetric under changing
signs of both $\lambda$'s in this spin polarized case. In Fig. \ref{cuts},
$\Delta_{12}/W$ is plotted as a function of $t_2$ for three values of
$\lambda_1=\lambda_2$. We see that at negative values of $t_2$ a lower
spin-orbit coupling is needed; while for positive values of $t_2$ higher values
of spin-orbit coupling would maximize the bandgap to bandwidth ratio.

\begin{figure}
\begin{center}
\includegraphics[width=2.95in]{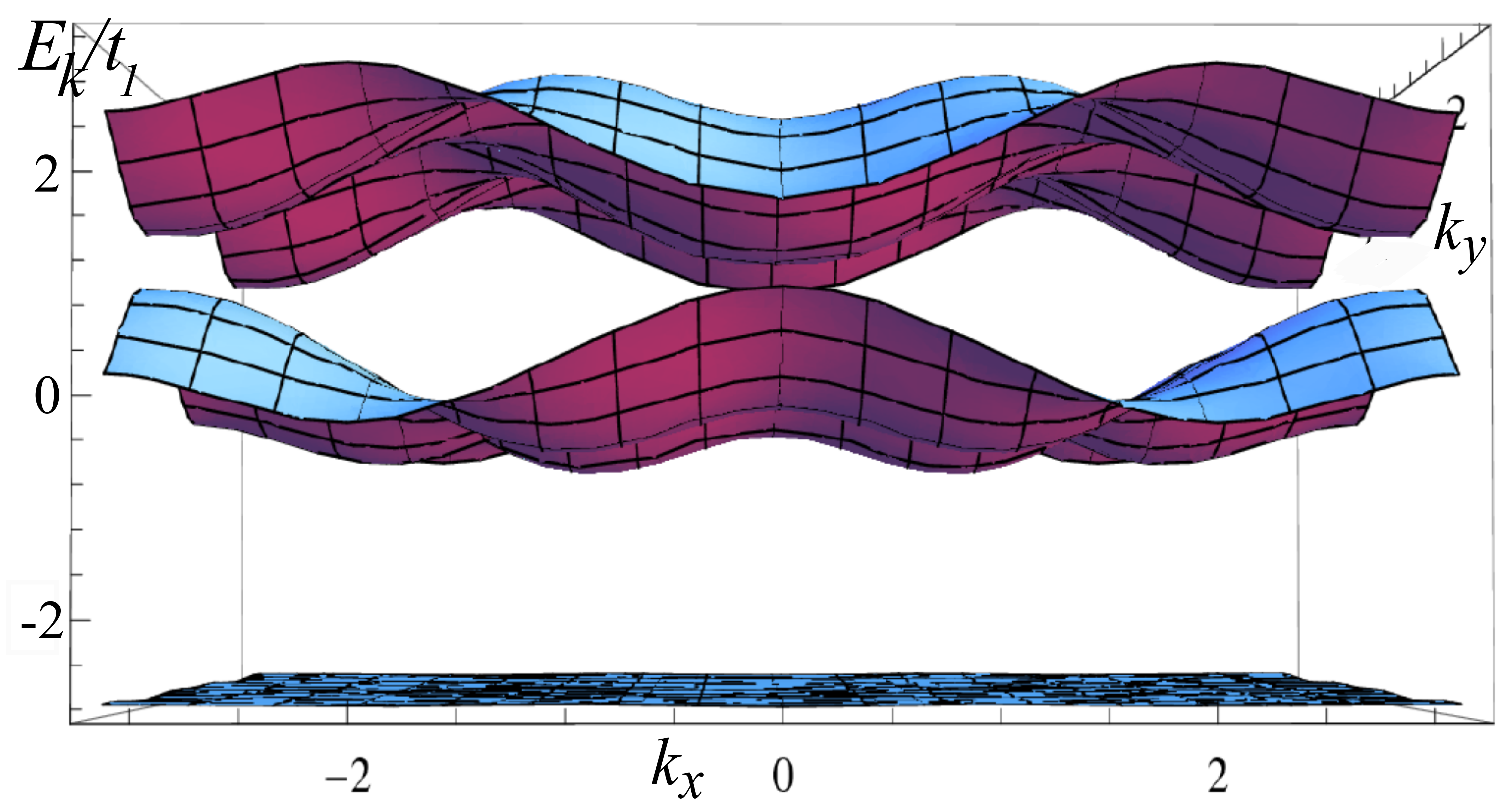}
\end{center}
\caption{(Color online) A very flat lowest band --- well-separated from the two
higher bands --- is obtained with the parameters $t_2=-0.3$, $\lambda_1=0.28$
and $\lambda_2=0.2$ (Case 1). The bandgap to bandwidth ratios are high:
$\Delta_{12}/W=52$ and $\Delta_{13}/W=99$ respectively.} 
\label{disp}
\end{figure}

\begin{figure}
\begin{center}
\includegraphics[width=2.8in]{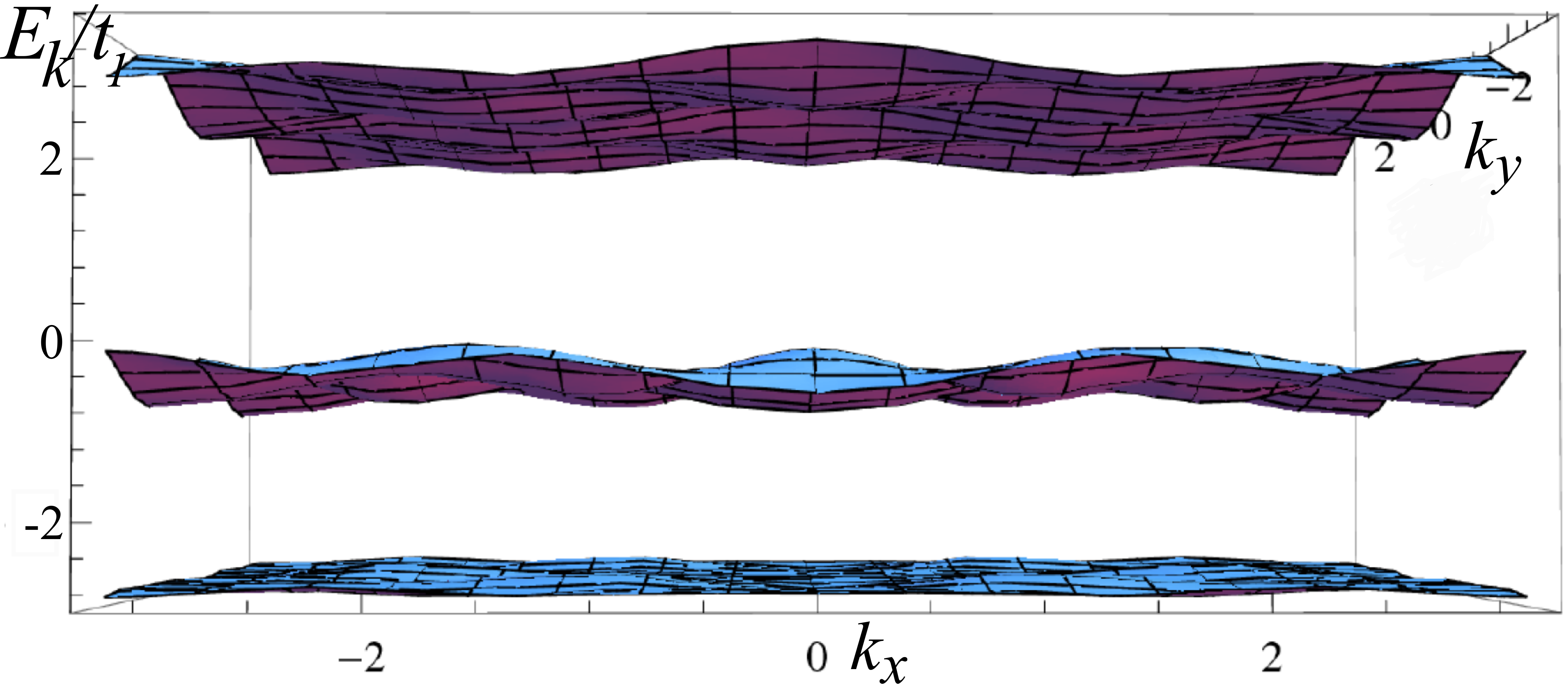}
\end{center}
\caption{(Color online) Three fairly flat bands are mutually well-separated
with bandgap to bandwidth ratios of $\Delta_{12}/W=8.7$ and $\Delta_{12}/W=24$.
Parameters used are $t_2=-0.3, \lambda_1=0.6$ and $\lambda_2=0$ (Case 2).}
\label{disp2}
\end{figure}

We present two examples where  $\Delta_{12}/W$ and $\Delta_{13}/W$ reach high
values at $t_2=-0.3$. In Case 1, setting $t_2=-0.3$, $\lambda_1=0.28$
and $\lambda_2=0.2$, we obtain a very flat lowest band separated from the two
higher bands by a large gap (see Fig. \ref{disp}).  The values of
$\Delta_{12}/W$ and $\Delta_{13}/W$ are $52$ and $99$ respectively.
In another example, all three bands are fairly flat (particularly the lowest
one) and mutually well-separated. The parameters used are $t_2=-0.3,
\lambda_1=0.6$ and $\lambda_2=0$. In Case 2, we obtain $\Delta_{12}/W=8.7$ and
$\Delta_{13}/W=24$ respectively (see Fig. \ref{disp2}). 

Calculating the Chern number $c$ of the lowest flat band in these two cases, we
find it is 1. This is expected as slowly turning off $t_2$ and
$\lambda_2$ does not close the bandgap --- and we have previously seen
that in the absence of next nearest-neighbor hopping, the lowest band always
has unit Chern number. When $\Delta\gg U\gg W$ is satisfied, partial filling of this flat band would favor the FQH state.  

\begin{figure}
\begin{center}
\includegraphics[width=3.33in]{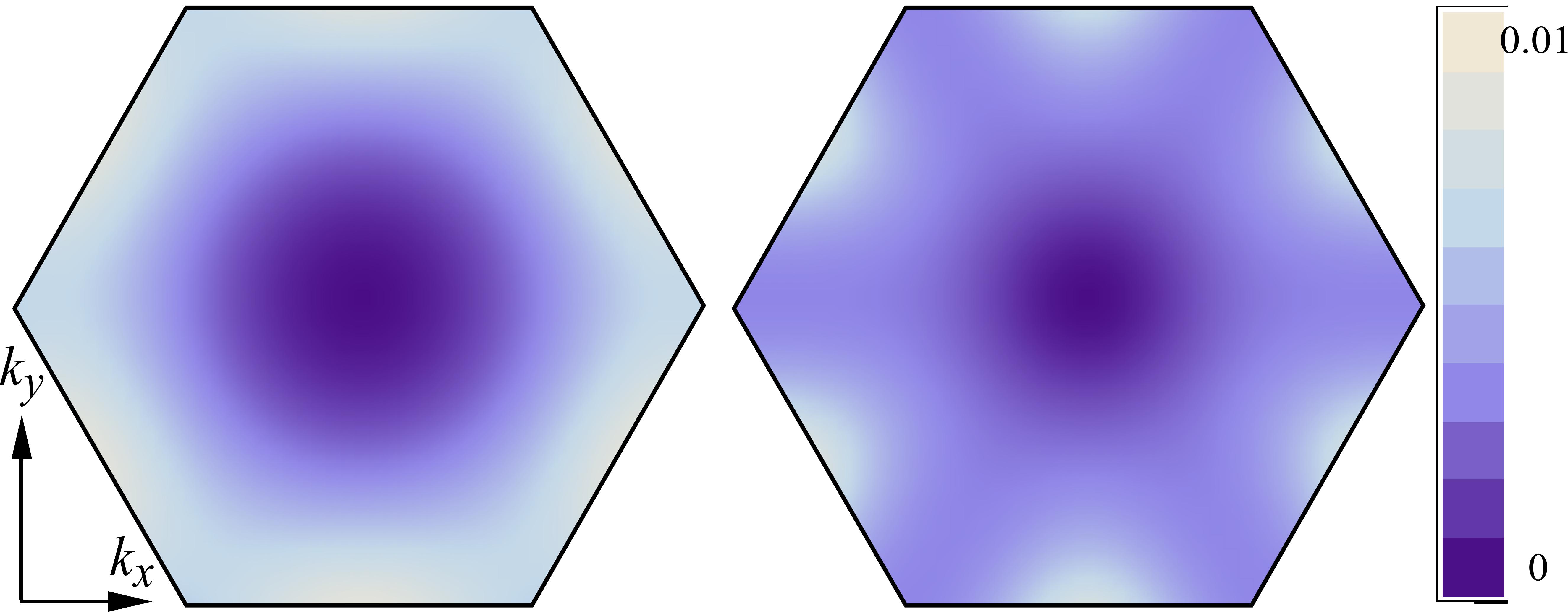}
\end{center}
\caption{(Color online) Distribution of the field strength $F_{12}(k)$ (Eq. \ref{fs}) in the Brillouin zone for the flat bands in Cases 1 and 2 discussed above. They do not contain sharp features --- especially Case 1 with the flatter band --- hence the magnetic length scale remains on the order of the lattice constant $a$.}
\label{berry}
\end{figure}

The distribution of the field strength $F_{12}(k)$ in the Brillouin zone is
plotted in Fig. \ref{berry}. We observe there are no singularities or very
sharp features but $F_{12}(k)$ varies fairly smoothly especially in the first
case with the flatter band. The presence of singularities --- e.g. localized at
the Dirac point --- would have signalled a new (and much larger) length scale
in the system. In our case, both the magnetic length scale (arising from
spin-orbit interactions) and the variation of field strength $F_{12}(k)$ are on
the order of the lattice constant $a$. 

Thus the interaction energy scale is generated from the lattice constant $a$:
$U\sim e^2/\eps a$ where $\epsilon$ is the dielectric constant. If the flat band is 1/3 filled, analogous to results from FQH states in semiconductor-based systems\cite{WST8876} the gap
for this $\nu=1/3$ state is here roughly $0.09 e^2/\eps a=500$K. (We choose $\eps=3$ and $a=10\AA$; $a$ is defined as the square root of the
unit cell area). As the interaction energy scale is a hundred times larger than in semiconductors, we may 
see the FQH effect at room temperature. As band gaps $\Delta_{12}$ and
$\Delta_{13}$ are easily much higher than room temperature, a fully filled band
could give the integer quantum Hall state at unusually high temperatures too.


\emph{Materials realization}
We see that a suitable combination of geometric frustration, ferromagnetism and
spin-orbit interactions can give rise to nearly flat bands with a large bandgap
and non-zero Chern number.  The ferromagnetism can arise from an external
magnetic field, a ferromagnetic substrate for a thin film sample, and/or
exchange effects.  If the flat band is close to the Fermi energy, partial
filling of the flat band can be controlled by doping and give rise to FQH
states at high temperatures.

The choice of parameters in our calculations is based on known values of
spin-orbit coupling. For example, in Herbertsmithite (a common copper-based 2D
kagome lattice) the spin-orbit interaction is $8\%$ of the kinetic
energy.\cite{ZNT0805} Other compounds with 4d or 5d orbitals (instead of 3d as
in Cu) may experience a larger spin-orbit interaction. For instance, the
strength of spin-orbit coupling in Iridium-based kagome compounds can be on the
order of magnitude of the kinetic energy, 
hence the substitution with 4d or 5d atoms in metallic kagome
lattices could result in hopping parameters similar to the ones used in our
work. Alternatively, making thin
films of frustrated lattices with 4d or 5d atoms may lead to a flat
band with strong spin-orbit coupling, where exchange effects in this flat band could cause ferromagnetism.

Most existing kagome compounds are Cu-based insulators.  
Some 2D kagome lattices show metallic behavior, for instance
$Fe_3Sn_2$.\cite{FDW0902,KFW0989}
This material shows ferromagnetism along the $c$-axis above 60K and in the
kagome plane below this temperature.  Also, as spin-orbit interactions can be
simulated in cold atom systems,\cite{LCG0928,GSN1009} it is possible to realize
our hopping model in such systems as well. This would provide a method to
obtain FQH states in cold atom systems. 

In short, flat bands with non-zero Chern number arise in the examples we have
given and in other geometrically frustrated systems with suitable levels of
spin orbit interaction. By partially filling these bands e.g. via doping, one
can expect the emergence of FQH states at high temperatures. While the
identification of exact material candidates remains open, our work indicates
intriguing directions for synthesis and development.

After completion of this paper, we learned that T. Neupert et. al\cite{Chamon2}
also discussed the possibility of the FQH effect in interacting two-band
lattice systems, while K. Sun et. al\cite{sarma} found flat bands with non-zero
Chern numbers on various lattices (including the kagome lattice) after
including some complex hopping. 

This research is supported by  NSF Grant No.  DMR-1005541.


%
\end{document}